\documentclass[11pt]{article}
\def\D{\Delta}
\def\d{\delta}
\def\L{\Lambda}
\def\l{\lambda}
\def\S{\Sigma}
\def\G{\Gamma}
\def\g{\gamma}
\def\e{\epsilon}
\def\s{\sigma}
\def\S{\Sigma}
\def\o{\omega}
\def\O{\Omega}
\def\i{\iota}
\def\a{\alpha}
\def\b{\beta}

\def\m{\mu}
\def\n{\nu}
\def\r{\rho}
\def\s{\sigma}
\def\p{\pi}

\def\f{\phi}
\def\F{\Phi}
\def\th{\theta}
\def\t{\tau}
\def\e{\epsilon}

\def\vf{\varphi}

\def\cd{{\cal D}}

\def\cl{{\cal L}}

\def\pa{\partial}

\def\det{\textrm{det}}
\def\const{\textrm{const.}}
\def\iu{\textrm{i}}
\def\R{{\bf R}}

\def\u{{\bf u}}
\def\v{{\bf v}}
\def\bv{{\Big |}}

\newcommand{\be}{\begin{equation}}
\newcommand{\ee}{\end{equation}}
\newcommand{\bea}{\begin{eqnarray}}
\newcommand{\eea}{\end{eqnarray}}

\begin{document}

\begin{center}
\bf{\Large  Finiteness of quantum gravity with matter on a PL spacetime}
\end{center}

\bigskip
\begin{center}
{\bf A. Mikovi\'c} \footnote{Member of  COPELABS.} \\
Departamento de Inform\'atica e Sistemas de Informac\~ao \\
Universidade Lus\'ofona\\
Av. do Campo Grande, 376, 1749-024 Lisboa, Portugal\\
and\\
Mathematical Physics Group of the University of Lisbon\\
Campo Grande, Edificio C6, 1749-016 Lisboa, Portugal
\end{center}

\centerline{E-mail: amikovic@ulusofona.pt}

\bigskip
\bigskip
\begin{quotation}
\noindent\small{We study the convergence of the path integral for General Relativity with matter on a picewise linear (PL) spacetime that corresponds to a triangulation of a smooth manifold by using a path-integral measure that renders the pure gravity path integral finite. This measure depends on a parameter $p$, and in the case when the matter content is just scalar fields, we show that the path integral is absolutely convergent for $p>0,5$ and not more than 2 scalar fields. In the case of Yang-Mills fields, we show that the path integral is absolutely convergent for the $U(1)$ group and $p>0,5$. In the case of Dirac fermions, we show that the path integral is absolutely convergent for any number of fermions and a sufficiently large $p$. When the matter content is given by scalars, Yang-Mills fields and fermions, as in the case of the Standard Model, we show that the path integral is absolutely convergent for $p > 46,5$. Hence one can construct a finite quantum gravity theory on a PL spacetime such that the classical limit is General Relativity coupled to the Standard Model.
 }\end{quotation}

\newpage
\section{Introduction}

A minimalist version of the problem of constructing a quantum gravity (QG) theory would be the following: given a 4-dimensional spacetime $M$ of the topology $\S\times I$, where $\S$ is a smooth 3-manifold and $I$ an interval from $\R$, construct a quantum theory such that its classical limit is given by the General Relativity (GR) coupled to the Standard Model (SM) on $M$, see \cite{I}. The simplest approach is just applying the quantum field theory (QFT) formalism to GR plus the SM, but as it is well known, such a QFT theory is perturbatively non-renormalizable, see \cite{S,V}, and hence non-physical, because it requires a knowledge of infintely many coupling constants. A formalism of non-perturbative renormalisability, called asymptotic safety, has been proposed by Weinberg in 1977, which could be applied to gravity \cite{W1,W2},  so that only finitely many coupling constants will be physical. It is a promissing idea to solve the problem, since it works in some special cases \cite{asqg}, but the main drawback is that there is no proof for the general case. 

The most celebrated proposal for a QG theory so far is the string theory, \cite{GSW,Pb}, which circumvents the non-renormalizability of GR by constructing a finite quantum pertubation  theory. However, this is achieved by introducing extra spatial dimensions and many more fields than what we have in GR and SM, which is the main drawback of string theory since a solution which corresponds to the SM at low energies still has not been found.

Another notable proposal is Loop Quantum Gravity (LQG), see \cite{R,T,Pz}. LQG has two formulations: the canonical quantization formulation \cite{T} and the path integral (PI) formulation \cite{R,Pz}. In the canonical quantization formulation, the main problem is solving the Wheeler-de Witt equation and obtaining solutions that have a semi-classical (SC) expansion which would correspond to the perturbative QFT expansion. In the path integral formulation, which is known as the spin foam (SF) models, one must use a piecewise linear (PL) spacetime $T(M)$ which corresponds to a triangulation of $M$ in order to define a SF model. Although a SF state sum can be made finite by an appropriate choice of the weights \cite{MV2}, the main problem is that it is not known how to define the smooth-manifold limit of a spin foam model. Another major problem is that it is very difficult to couple fermions in a SF model, because a fermion on $T(M)$ couples to gravity via vierbeins while in a SF model one only has the 2-forms which correspond to the wedge products of the vierbein 1-forms.

Note that a SF model represents a generalization of the Regge path integral, and also note that there is no problem in coupling fermions in the Regge calculus, see \cite{H, MVb}. In the Regge approach to QG, see \cite{H}, there is also the problem of the smooth-manifold limit, but there is a promissing idea to solve it. One can use the Wilsonian approach, i.e. to look for the critical points of the Regge path integral as a function of the GR couplings (the Newton constant $G_N$ and the cosmological constant), see \cite{H}. Near a critical point, the correlation length becomes very large, so that the effective theory becomes a field theory on a smooth manifold. However, the QFT perturbative expansion corresponds to small $G_N$ values, and this requires a nonperturbative evaluation of the path integral. Such a nonperturbative evaluation of the path integral would be a way to prove the main assumption of the asymptotic safety approach, but so far one can only do the numerical simulations, see \cite{cdt}. 

A promissing new framework, which is based on the Regge calculus, was proposed in \cite{M}, see also \cite{MV, MVr, M1, MVb}.  The idea is to use  the triangulations with a large number of simplicial cells, but without taking the smooth limit, since it is assumed that the short-distance structure of the spacetime is given by the PL manifold $T(M)$. Then the PL manifold appears to be smooth at the distances much larger than the average edge length, and this can be clearly seen by using the effective action formalism from QFT. Another novelty with respect to the standard quantum Regge calculus, is utilization of a non-trivial path integral measure, whose form is constrained by the requirements of the finitenness of the path integral and the existence of the correct classical limit for the effective action, see \cite{M,M1}. 

One can then choose a simple form of the measure, see eq. (\ref{plm}),
which depends on a length parameter $L_0$ and a dimensionless parameter $p$, such that $p > 1/2$ insures the absolute convergence of the GR path integral \cite{M1}. However, there has been no investigation of the finitennes of the PLQG path integral when matter is included, except in the case of a minisuperspace model \cite{M2}. In this paper we are going to explore the convergence of the path integral of GR coupled to scalar, Yang-Mills and spinor fields on $T(M)$ with the measure (\ref{plm}). In section 2 we briefly describe the PLQG formalism for GR. In section 3 we explore the case when one couples the scalar fields, while in section 4 we study the case of Yang-Mills fields. In section 5 we study the case of Dirac spinors, while in section 6 we investigate the convergence of the PLQG path integral when we have the matter fields of the Standard Model. In section 7 we present our conclussions.

\section{GR path integral}

Let $M$ be a smooth 4-manifold and $\S$ a three-dimensional submanifold. We will restrict the topology of $M$ in the following way: when $M$ has no 
boundary, $M$ is equivalent to the result of gluing two 4-manifolds $M_1$ and $M_2$ to the boundaries of the manifold $\S \times I$, where $I$ is an interval from $\R$. Therefore
$$M = M_1 \cup \left(\S\times I \right) \cup M_2 =  M_1 \cup M_{12} \cup M_2\,,$$
such that  $\pa M_1 = \pa M_2 = \S$.

The topology of $M_1$ and $M_2$ can be arbitrary, but in the context of quantum cosmology, when $M$ has a single boundary, then
$M = M_1$ or $M=M_1 \cup M_{12}$. In this case one can choose $M_1$ to have the simplest possible topology.  For example, when $\S = S^3$, one can take $M_1 = S^4$, see \cite{M2}. In the case when we are interested in the effective action, we take $M = M_{12} = \S \times I$, see \cite{MVr}.

Let $T(M)$ be a PL manifold corresponding to a regular\footnote{The dual one simplex is a connected 5-valent graph.} triangulation of $M$. A metric on $T(M)$ is given by a set of edge lengths $L_\e$ such that  $L_\e^2 \in \R$.
We can then choose  $L_\e \in \R_+$ or $L_\e \in i\R_+$, which corresponds to having a spacelike or a timelike distance between the vertices of $T(M)$.

We also require that $|L_\e| >0$ for a compact $M$, i.e. we exclude the light-like edges and the zero-vector edges. For a noncompact $M$, we take  $|L_\e| > 0$ for the edges in a $T(B_3 \times I) \subset T(\S\times I)$, where $B_3$ is a 3-dimensional ball in $\S$. For the edges outside of $T(B_3 \times I)$ we take $L_\e = 0$ as zero vectors. In the case of $M_1$ or $M_2$, we take $|L_\e| > 0$ for the edges in a $T(B_4) \subset T(M_i)$.

Given a set of the edge lengths, one can define a metric on the PL manifold $T(M)$  such that the PL metric in each 4-simplex $\s$ of $T(M)$ is flat. This can be done by using the Cayley-Menger (CM) metric, see \cite{H, MVb}
\be G_{\m\n}(\s) = L_{0\m}^2 + L_{0\n}^2 - L_{\m\n}^2 \,,\label{cmm}\ee
where the five vertices of $\s$ are labeled as $0,1,2,3,4$ and $\m,\n = 1,2,3,4$. 

Although the CM metric is flat in a four-simplex, it is not dimensionless and hence it is not diffeomorphic to 
$$g_{\m\n} = diag(\pm 1, \pm 1, \pm 1, \pm 1) \,.$$ 
This can be remedied by defining a new PL metric \cite{MV}
\be g_{\m\n}(\s) = {G_{\m\n}(\s) \over |L_{0\m}||L_{0\n}|} \,.\label{eplm}\ee

In order to insure that we are dealing with a PL Lorentzian manifold, one has to choose the edge lengths such that $g_{\m\n}$ is difeomorphic in each $\s$ to
$$ \eta_{\m\n} = diag(-1,1,1,1)\,. $$
This can be done by embedding $T(M)$ into $\R^5$ with a 5-dimensional Minkowski metric, see \cite{M2}.

The Einstein-Hilbert (EH) action on $M$ is given by
\be S_{EH} = \int_M \sqrt{|\det g|}\, R(g) \, d^4 x  \,,\label{eh}\ee
where $R(g)$ is the scalar curvature associated to a metric $g$. On $T(M)$ the EH action becomes the Regge action 
\be S_R (L) = \sum_{\D\in T(M)} A_\D (L)\, \d_\D (L) \,,\label{era}\ee
when the edge lengths correspond to a Eucledean PL geometry. In (\ref{era}), $A_\D$ denotes the area of a triangle $\D$, while  the deficit angle $\d_\D$ is given by
\be \d_\D = 2\pi - \sum_{\s \supset \D} \th_\D^{(\s)} \,,\ee
where a dihedral angle $\th_\D^{(\s)}$ is defined as the angle between the 4-vector normals associated to the two tetrahedrons that share the triangle $\D$. 

In the case of a Lorentzian geometry, a dihedral angle can take complex values, so that it is necessary to modify the formula (\ref{era}) such that the Regge action takes only the real values. This can be seen from the formula
\be \sin \th_\D^{(\s)} = \frac{4}{3} {v_\D v_\s \over v_\t v_{\t'}} \,, \ee
where $v_s$ is the volume $V_s$ of the simplex $s$ ($V_s \ge 0$ by definition) if the CM determinant is positive, while $v_s = \iu V_s$  if the CM determinant is negative. Consequently,  the value of $\sin \th_\D^{(\s)}$ may not be restricted to the interval $[-1,1]$, but $\sin \th_\D^{(\s)} \in \bf R$ or $\sin \th_\D^{(\s)} \in \iu{\bf R}$. This implies that the Lorentzian dihedral angles can take the complex values, see \cite{cdt, M1}.

Consequently, the Regge action (\ref{era}) will give a complex number when the spacelike triangles are present. One can modify the Regge action as
\be  S^*_R(L) = Re\left(\sum_{\D(s)} A_{\D(s)} \, \frac{1}{\iu}\,\d_{\D(s)} \right) + \sum_{\D(t)} A_{\D(t)} \,\d_{\D(t)}  \,,\label{lra} \ee
where $\D(s)$ denotes a spacelike triangle, while $\D(t)$ denotes a timelike triangle, so that it is always real and corresponds to the Einstein-Hilbert action on $T(\S\times I)$, see \cite{cdt,M2,M1}.

Consequently
$$ Z(T(M)) = \int_D \prod_{\e=1}^N dL_{\e} \, \m(L) \, e^{\iu S^*_R(L)/l_P^2} \,, $$
where $dL_\e = d|L_\e|$ and $\m(L)$ is a mesure that ensures the finiteness and gives the effective action with a correct semiclassical expansion, see \cite{M,M1}. The integration region $D$ is a subset of ${\bf R}_+^N$, consistent with a choice of spacelike and timelike edges.

It was shown in \cite{M1} that $Z(T(M))$ is convergent for the measure
\be \m(L) = e^{-V(M)/L_0^4}\prod_{\e=1}^N \left(1+ {|L_\e|^2 \over l_0^2} \right)^{-p} \,,\label{plm}\ee
where 
\be p > 1/2 \,.\label{pgp}\ee 

The convergence of $Z$ for large $|L_\e|$ is guaranteed by the $(1+|L_\e|^2 /l_0^2)^{-p}$ factors, since the exponentially damped measure 
\be \m_0 (L) = e^{-V_4(L)/L_0^4 } \,,\ee
is not sufficient to guarantee the convergence, due to the existence of degenerate configurations, i.e. regions in $D$ such that $V_4(L) \to 0$ while some or all $|L_\e| \to \infty$, see \cite{M1}.

The measure (\ref{plm}) also insures that the corresponding effective action (EA) can be approximated by a smooth-manifold EA. This happens when the number of the edges is large, i.e. $N\to\infty$, and $L_\e$ are sufficiently small, i.e. $|L_\e| \le \g l_0 / N$, where $\g$ is a constant, see \cite{M1}.

The smooth-manifold effective action is then given by the QFT effective action for GR with a momentum cutoff $K$ determined by the average edge length, so that
\be\G(L) \approx \G_K [g_{\m\n}(x)] \,,\label{smea}\ee
when $|L_\e| \to 0$ and $N\to \infty$, where $K=2\pi /\bar L$, see \cite{MVb}. The smooth metric in (\ref{smea})  obeys
$$ g_{\m\n}(x) \approx  g_{\m\n}^{(\s)}(L)  \,, \quad \textrm{for} \,\, x\in\s \,.$$
Furthermore, if $|L_\e|$ are small, but still much larger than $l_P$, then $\G_K(g)$ is given by the QFT perturbative expansion
$$\G_K (g) = S_{EH} (g) + l_P^2\, \G_K^{(1)}(g) + l_P^4 \,\G_K^{(2)}(g) + \cdots \,, $$
where $\G_K^{(n)}$ is the $n$-loop QFT effective action for GR, see \cite{M}.

Note that one can also add the cosmological constant (CC) term to the Regge action, so that
$$ S_R^*(L) \to S_R^*(L) + \L_c V_4 (L) \, . $$
Then the condition for the semi-classical expansion of the EA, $|L_\e| \gg l_P$ and $L_0 \gg l_P$, is substituted by $|L_\e| \gg l_P$ and
\be L_0 \gg \sqrt{l_P L_c} \,,\label{lcr}\ee 
where $|\L_c| = 1/L_c^2$ \cite{MV}.

\section{Scalar fields}

When the matter fields $\F$ on $T(M)$ are present, they are described by a finitely many degrees of freedom (DOF), since the uncountably many values of $\f(x)$, $x\in M$, are replaced by finitely many values $\F(\p)$, where $\p$ are the vertices of $T(M)$. One can also use as the matter DOF the values of $\F(v)$, where $v$ are the dual vertices\footnote{Given a PL manifold $T(M)$, the dual PL manifold $T^*(M)$ has one vertex $v$ inside every 4-simplex $\s$ of $T(M)$ and a dual edge $l$ intersects exactly one tetrahedron $\t$ of $T(M)$.}. 

The corresponding effective action has the same properties as in the pure gravity case when $N\to\infty$ and $|L_\e|\to 0$, i.e.
$$\G(L,\F) \approx \G_K [g_{\m\n}(x), \f(x)] \,,$$
where $\G_K$ is the QFT effective action for the cutoff $K = 2\pi/\bar L$ and
$$ g_{\m\n}(x) \approx g^{(\s)}_{\m\n}(L) \,, \quad \f (x) \approx \F(v)  \quad  \textrm{for}\,\, x\in\s  \,\, \textrm{and}\,\, v\in\s \,.$$

We also have
$$\G_K (g,\f) = S_{EH} (g) + S_m (g,\f) +  l_P^2\, \G_K^{(1)}(g,\f) + l_P^4 \,\G_K^{(2)}(g,\f) + \cdots \,, $$
for $|L_\e| \gg l_P$ and small $\f$, where $\G_K^{(n)}$ is the $n$-loop QFT effective action for GR coupled to matter, see \cite{MV}. However, the non-perturbative effective action $\G(L,\F)$ will be only defined if the gravity plus matter path integral is finite. In this section we will study the case of scalar fields for the PI measure (\ref{plm}).

When the matter is added, one expects that the PI measure (\ref{plm}) will be sufficient, since the integrals over the discrete matter fields do not have the problem of the degenerate configurations. However, the convergence of the matter path integral
$$ Z_m (L) = \int_{{\bf R}^n} d^n \vf \, e^{\iu S_m(\vf, L)/\hbar} \,,$$
where $\vf$ is a matter field and $n$ is the number of 4-symplex cells in $T(M)$,
is still not guaranteed. 

This problem can be resolved by passing to a Eucledean geometry defined by the edge lengths
$$ \tilde L_\e = |L_\e| \,, $$
so that all the Eucledean edge lengths are positive real numbers. This is equvalent to a Wick rotation where $\tilde L_\e = L_\e$ if $\e$ is a spacelike edge and $\tilde L_\e = (-\iu)L_\e$, if $\e$ is a timelike edge.

Then we will consider the integral
\be \tilde Z_m ( \tilde L) = \int_{{\bf R}^n} d^n \vf \, e^{-\tilde S_m(\vf,  \tilde L)/\hbar} \,,\label{zm}\ee
where $\tilde S_m$ is the Euclidian matter action. Since $\tilde S_m (\vf, \tilde L)$ is a positive function of $\vf$, and 
$$ \tilde S_m (\vf, \tilde L) \to +\infty\,, \quad \textrm{for} \,\, |\vf| \to +\infty \,, $$
 then the integral $\tilde Z_m$ will be convergent. Hence we will define
\be Z_m(L) = \tilde Z_m (\tilde L)\bv_{\tilde L = w(L)} \,,\ee
where $w$ is the Wick rotation.

Therefore the gravity plus matter path integral will be given by
\be Z = \int_D d^N L \,\mu(L)\, e^{\iu S^*_R(L)/l_P^2}\, Z_m(L) \,,\label{gsfpi}\ee
and the convergence of the integral $Z$ will be studied. 

Consider a scalar massive field on $T(M)$. Then
\be \tilde S_m (\vf) = \frac{1}{2} \sum_{\s\in T(M)} V_\s (\tilde L) g_\s^{\m\n}(\tilde L) \D_\m \vf \D_\n \vf + \sum_{\p\in T(M)} U(\vf_\p) V_\p (\tilde L) \,,\label{sfa}\ee
where
$$U(\vf) =\frac{1}{2} m^2 \vf^2 + \l_0 \vf^4 \,,$$
and $V_\p$ is the 4-volume of the dual cell around the vertex $\p$ of $T(M)$. The PL derivatives $\D_\m$ are defined as
$$ \D_\m \f (\s) = \frac{1}{5}\sum_{\p_0 \in\s} {\vf (\p_\m ) - \vf (\p_0) \over \tilde L_{0\m}} \,, $$
where $\p_0$ and $\p_\m$, $\m = 1,2,3,4$, are the verices of a 4-symplex $\s$.


The action (\ref{sfa}) can be written as
$$\tilde S_m  = \sum_{\p,\p'} K_{\p\p'}(\tilde L) \vf_\p \vf_{\p'} + m^2 \sum_\p V_\p (\tilde L) \vf^2_\p + \l_0 \sum_\p V_\p (\tilde L) \vf^4_\p  \,.$$

It will be important to notice that $K(L)$ and $V(L)$ are homogenious functions of degree 2 and 4, so that
$$ K(\l L) = \l^2 K(L) \,,\quad V(\l L) = \l^4 V(L) \,. $$

Let us now  introduce the $N$-dimensional spherical coordinates in $\R_+^N$
\be |L_\e| =r \, f_\e ( \th_1, \cdots, \th_{N-1}) \,,\ee
such that
\be r^2 = \sum_{\e =1}^N |L_\e|^2 \,,\label{rl}\ee
and 
\be \th = (\th_1,..., \th_{N-1}) \in \O_+ = \left[0, \frac{\pi}{2}\right]^{N-1} \,. \ee

Let us make the change of variables $ \xi_\p = r \vf_\p$. The scalar-field action can be then written as
$$ \tilde S_m (\vf,\tilde L) = \sum_{\p,\p'} k_{\p\p'}(\th) \xi_\p \xi_{\p'}  + m^2 r^2 \sum_\p v_\p (\th) \xi^2_\p + \l_0 \sum_\p v_\p (\th) \xi^4_\p \,,$$
where
$$ K(\tilde L) = r^2 k(\th) \,, \quad V(\tilde L) =r^4 v(\th) \,.$$

When $m=0$, then one can show that
\be \tilde Z_0 ( \tilde L) = r^{-n} F_n (\th) \,,\label{ezm}\ee
and consequently
\be Z_0 ( L) = r^{-n} F_n (\th) \,.\label{lzm}\ee

The relation (\ref{ezm}) follows from 
$$\tilde Z_0 (\tilde L) = r^{-n} \int_{\R^n} d^n \xi \, e^{- s(\th,\xi)} = r^{-n}\,F_n (\th) \,,$$
where 
$$s(\th,\xi) = \tilde S_m (\tilde L, \vf)\bv_{m=0} = \sum_{\p,\p'} k_{\p\p'}(\th) \xi_\p \xi_{\p'}  + \l_0 \sum_\p v_\p (\th) \xi^4_\p \,.$$ 

The relation (\ref{lzm}) follows from the fact that the points $(\tilde L_1, \cdots,\tilde L_N)$ and $(L_1, \cdots, L_N)$ have the same spherical coordinates.

Since the integral which defines $F_n(\th)$ is finite and positive for any $\th\in \O_+$,  therefore $F_n(\th)$ is a bounded positive function on $\O_+$. Since $\O_+$ is a compact set, then
$$0 < F_n(\th) \le C_n \,.$$

Let us now study the convergence of the total path integral (\ref{gsfpi}). Let us divide the integration region $D$ such that
$$D = D_1 \cup D_2 \,, $$
where
$$ D_1 = D \cap B(R) \,, \,\, D_2 = D \cap \bar B(R) \,,$$
where $B(R)$ is a  ball $r \le R$ and $\bar B(R) = \R_+^N \setminus B(R)$.

Therefore
$$ Z = Z_1  + Z_2\,, $$
where
$$Z_k = \int_{D_k} d^N L \,\mu(L)\, e^{\iu S^*_R(L)/l_P^2}\,  Z_0(L) \,.$$

Consequently
$$|Z| \le |Z_1|  + |Z_2| \,,$$
and we have 
$$ |Z_1| = \bv \int_{D_1} d^N L \,\mu(L)\, e^{\iu S^*_R(L)/l_P^2}\,  Z_0(L) \bv \le \int_{D_1} d^N L \,\mu(L)\, |Z_0(L)|\,. $$

Since $0< \m(L) \le 1$, then 
\be |Z_1| \le \int_{D_1} d^N L\, |Z_0(L)| = \const\int_0^R r^{N-1-n} d r \int_{\O} d^{N-1}\th J_N(\th) F_n (\th)\,,\label{ri}
\ee
where $\O$ is the set of angles for the points in $D$.

The radial integral in (\ref{ri}) will be convergent for 
$$N  > n \,,$$ 
or
\be N_1 > N_0 \,,\label{ccsf}\ee
which is satisfied for many regular triangulations. The angular integral is also finite, since $\O \subseteq \O_+$, and $J_N$ and $F_n$ are bounded functions  on $\O_+$. Hence $Z_1$ is absolutely convergent.

In the case of the integral $Z_2$ we have
$$ |Z_2| = \bv \int_{D_2} d^N L \,\mu(L)\, e^{\iu S^*_R(L)/l_P^2}\,  Z_0(L) \bv \le \int_{D_2} d^N L \,\mu(L)\, |Z_0(L)| \,,$$ 
so that
\bea |Z_2|&\le&  \int_{D_2} d^N L \, e^{-V_4(L)/L_0^4}\, \prod_{\e=1}^N {l_0^{2p}\over (l_0^2 + L_\e^2)^p}\, {F_n (\th)\over r^n}\nonumber \\
&\le&  \int_{D_2} d^N L \, e^{-V_4(L)/L_0^4}\, \prod_{\e=1}^N {l_0^{2p}\over (l_0^2 + L_\e^2)^p}\, {C_n \over R^n} \,,\eea
since $F_n(\th) \le C_n$ for $\th\in\O_D \subseteq\O_N^+$ and $r \ge R$. Then
\bea |Z_2| &\le& {C_n \over R^n}\int_{D_2} d^N L  \prod_{\e=1}^N {l_0^{2p}\over {l_0^2 +  L_\e^2}^p} \nonumber\\
& < &{C_n \over R^n}\int_{\R_+^N} d^N L  \prod_{\e=1}^N {l_0^{2p}\over( l_0^2 +  L_\e^2)^p} = {C_n \over R^n}\prod_{\e=1}^N \int_0^\infty {l_0^{2p} \,dL_\e\over (l_0^2 +  L_\e^2)^p} \,. \eea

Since $p > 1/2$, this implies that $Z_2$ is absolutely convergent and consequently $Z$ is convergent for a massless scalar field.


When the scalar field mass $m$ is different from zero, we have 
$$ 0 <  Z_{m > 0}(L) < Z_{m=0} (L) \,,$$
because
$$\tilde S_m (\tilde L, \vf)\bv_{m>0}  > \tilde S_m (\tilde L, \vf)\bv_{m=0}\,.$$

Therefore
\be |Z| \le \int_{D} d^N L \,\mu(L)\, |Z_m(L)| < \int_{D} d^N L \,\mu(L)\, |Z_0(L)| \,, \label{mzc}\ee
and since we have showed that the last integral in (\ref{mzc}) is convergent, this implies the absolute convergence of $Z$ in the case of a massive scalar field.

When we have $k$ scalar fields, the absolute convergence condition changes to
\be N_1 > k N_0 \,.\label{ccsf}\ee

Note that a regular triangulation has a dual one-simplex which is a five-valent graph, so that $5N_0^* = 2N_1^*$. One can conjecture that 
\be {N \over n} = {N_1 \over N_0} \ge {N_1^* \over N^*_0} = \frac{5}{2} \,,\label{4de}\ee
which can be checked in the case of regular  triangulations of $S^4$ with a small number of vertices. For example, it holds in the case of the triangulation with $N_0 =6$ and $N_1 = 15$ ($T_1(S^4)$ = 5-simplex graph), or for a triangulation with $N_0 = 12$ and $N_1 = 55$ ($T_1(S^4)$ = two 5-simplex graphs joined by 5 parallel edges). 

In an arbitrary spacetime dimension $d$, we will have
\be {N_1 \over N_0} \ge {N_1^* \over N^*_0} = \frac{d+1}{2} \,,\label{dde}\ee
which can be checked for regular triangulations of $S^d$ with a small number of vertices. For example, in the case $d=2$, the inequality (\ref{dde}) holds for the triangulation with $N_0 = 4$ and $N_1 = 6$ ($T_1(S^2)$ = tetrahedron graph) and for triangulations with $N_0 = 8$ and $N_1 = 18$ (two tetrahedron graphs joined by three parallel edges). 

Since $N_1/N_0 \ge 5/2$ for a regular triangulation, from (\ref{ccsf}) it follows that for $k \ge 3$ the convergence of the PI is not guaranteed.

\section{Yang-Mills fields}
Consider a gauge field $A_\m (x) = A_\m^I (x) T_I $ on $M$, where $T_I$ is a basis of the Lie algebra for a Lie group $G$. The Yang-Mills (YM) action is given by
$$ S_{YM} = -\frac{1}{4}\int_M d^4 x \sqrt{-g}\,  g^{\m\r}g^{\n\i}\, Tr\left( F_{\m\n} F_{\r\i}\right) \,.$$

On $T(M)$ it is natural to use the variables
$$ A_\e = \frac{1}{|L_\e|}\int_\e A_\m dx^\m \,,$$
so that one can define the edge holonomies
\be U_\e = \exp(\iu |L_\e|A_\e ) \,.\label{wl}\ee

The edge holonomies transform under the gauge transformations as
$$U_{\e} \to g_\p \, U_{\e}\, g_{\p'}^{-1} \,,$$
where $g_\p \in G$, so that the field strengths can be associated to the triangles and one has
$$ F_\D = F_{123} \approx Im\, {U_{12}U_{23}U_{31} - Id \over A_{123}} = Im\, {U_{\D} - Id\over A_\D} $$
for $A_\D \to 0$.

One can then write a gauge invariant action
\be S_{YM}(U,L) = -\frac{1}{40}\sum_{\s\in T(M)} V_\s (L) \,\sum_{\e\in\s} Tr\left( F_{\D} F_{\D'}\right) \,, \label{ymtm}\ee
where $\D,\D' \in \s$ and $\D \cap \D' = \e$.


The action (\ref{ymtm}) is a function of the Wilson loop variables $U_\D$, which are qauge invariant, and this is the reason why they are used in QCD\footnote{Since QCD is formulated on a flat spacetime, one uses the plaquetes instead of triangles \cite{W}.}.  However, for the purposes of studying the finitenness of the PLQG path integral, it is more convinient to use the local gauge fields $A_\m (x)$, which can be defined by their values at the vertices of $T(M)$ as in the case of the scalar field. 

For a point $x\in\s$ we have
\be \pa_\m A_\n (x) \to  \D_\m A_\n (\s) =\frac{1}{5}\sum_{\p_0 \in\s}  {A_\n (\p_\m ) - A_\n (\p_0) \over \tilde L_{0\m}} \,, \label{dpd}\ee
so that one can define
\be F_{\m\n}(\s) =  \D_{[\m} A_{\n]} +  \frac{g_0}{5}\sum_{\p \in\s} [A_\m (\p), A_\n (\p) ] \,, \label{dfs}\ee
which can be associated to the field strength
$$ F_{\m\n}(x) = \pa_{[\m} A_{\n]}(x) + g_0 [A_\m (x), A_\n (x) ] \,,$$
where $g_0$ is the YM coupling constant.

As far as the gauge invariance
$$ A_\m(x) \to g(x) (A_\m( x) + \pa_\m ) g^{-1}(x) \,, $$
is concerned, one can define it on $T(M)$ as
$$ A_\m(\p) \to g_\p (A_\m(\p) + \D_\m ) g_\p^{-1} $$
where
$$ \D_\m  f(\p) = {f(\p_\m) - f(\p) \over \tilde L_{0\m}} \,.$$

The path integral for the YM action on $M$ can be defined by fixing a gauge, for example the Lorentz gauge
$$\pa^\m A_\m = 0  \,.$$
The path integral on $T(M)$ will be then given by a discretization of the formal expression \cite{FP}
\be Z_{YM} = \int \cd A\, \cd c \,\cd \bar c \, e^{\iu (S_{YM}(A) + S_{gf}(A) + S_{gh}(A,\bar c , c) )/\hbar}\,, \label{ympi}\ee
where $\bar c$ and $c$ are the Faddeev-Popov ghost fields and
\bea S_{YM} &=& -{1\over 4}\int_M d^4 x \, \sqrt{g} \,g^{\m\r}g^{\n\s}Tr\left(F_{\m\n} F_{\r\s}\right) \,, \label{sym}\\
 S_{gf} &=& -{1\over 2}\int_M d^4 x \, \sqrt{g} \,g^{\m\n}g^{\r\s}Tr\left(\pa_\m A_\n \, \pa_\r A_\s\right) \,,\label{sgf}\\
S_{gh} &=&    -\int_M d^4 x \,\sqrt{g}\,Tr\left( \pa^\m\bar c \, (\pa_\m c -\iu\, g_0 A_\m c) \right)   \,.\label{sgh}\eea

Although the PI (\ref{ympi}) corresponds to a fixed gauge for $A$, it is still gauge invariant, in the sense that it will give the same value in some other gauge. 

Hence we define on $T(M)$
\be \tilde Z_{YM} = \int \prod_{\p,\m,a} dA_\m^a (\p )\, \prod_{\p,a} d\bar c^a_\p \, d c^a_\p \, e^{- (\tilde S_{YM}(A) + \tilde S_{gf}(A) + \tilde S_{gh}(A,\bar c , c) )/\hbar}\,, \label{dympi}\ee
where now 
\bea\tilde S_{YM} &=& {1\over 4}\sum_\s V_\s (L) \,g^{\m\r}(\s)g^{\n\s}(\s) Tr\left(F_{\m\n}(\s) F_{\r\s}(\s)\right) \,, \label{eym}\\
 \tilde S_{gf} &=& {1\over 2}\sum_\s V_\s(L) \,g^{\m\n}(\s)g^{\r\s}(\s)Tr\left(\D_\m A_\n (\s) \, \D_\r A_\s (\s)\right) \,,\label{egf}\\
\tilde S_{gh} &=&   \sum_\s V_\s (L) \,g^{\m\n}(\s) \,Tr\left( \D_\m\bar c \, (\D_\n c -\iu\, g_0 A_\n  c) \right)   \,,\label{egh}\eea
are the Eucledean PL geometry versions of (\ref{sym}), (\ref{sgf}) and (\ref{sgh}), so that $\tilde S_{YM} \ge 0$ and $\tilde S_{gf} \ge 0$.

The action in the exponent of the integrand in (\ref{dympi}) can be rewritten as
$$ \tilde S_{YM } + \tilde S_{gf} = \sum_{p,q} K_{pq}(L) A_p A_{q} + g_0 \sum_{p,q,r} I_{pqr} (L) A_p A_{q} A_{r} \,, $$
$$ +\, g_0^2 \sum_{p,q,r,s} J_{pqrs} (L) A_p A_{q} A_{r} A_{s} \,,  $$
where $p=(\p, a)$, $\p\in T_0 (M)$ and $a = 1,2,..., |G|$, while
$$ \tilde S_{gh} =  \sum_{p,q} \tilde K_{pq}(L) \bar c_p c_{q} + g_0 \sum_{p,q,r} \tilde I_{pqr} (L) \bar c_p c_{q} A_{r} \,, $$
$$ +\, g_0^2 \sum_{p,q,r,s} \tilde J_{pqrs} (L) \bar c_p c_{q} A_{r} A_{s} \,. $$

We can integrate the Grassman variables $\bar c$ and $c$ in the path integral (\ref{dympi}), so that we obtain
\be \tilde Z_{YM} = \int \prod_{\p,\m,a} dA_\m^a (\p )\, \, e^{- \tilde S_{YM}(A) - \tilde S_{gf}(A) )/\hbar} \,\det\left( \tilde K + g_0 \tilde I A + g_0^2\tilde J AA \right)\,,\label{bdympi}\ee
where the matrix $\tilde K + g_0 \tilde I A + g_0^2\tilde J AA$ is given by
$$ \tilde K_{pq} + g_0 \sum_r \tilde I_{pqr} A_r + g_0^2\sum_{r,s}\tilde J_{pqrs} A_r A_s  \,.$$

Since
$$ K(\l L) = \l^2 K(L)\,, \quad I(\l L) = \l^3 I(L)\,, \quad J(\l L) = \l^4 J(L)\,, $$
and the same holds for $\tilde K$, $\tilde J$ and $\tilde L$, then
the change of variables
$$\xi_p = r A_p \,, \quad \forall p\in T_0 (M) \times I_G \,,$$
where $I_G =\{1,2,...,|G|\} $ and $r$ is the radial coordinate (\ref{rl}), gives
\be \tilde S_{YM} +\tilde S_{gf} = \sum_{p,q} k_{pq}(\th) \xi_p \xi_{q} + g_0 \sum_{p,q,r} i_{pqr} (\th) \xi_p \xi_{q} \xi_{r} + g_0^2 \sum_{p,q,r,s} j_{pqrs} (\th) \xi_p \xi_{q} \xi_{r} \xi_{s} \,,  \ee
where
$$ K = r^2 k(\th) \,,\quad I = r^3 i(\th) \,,\quad J = r^4 j(\th) \,.$$

Consequently we have
$$ \tilde Z_{YM} ( \tilde L) = r^{-4n|G|}\int_{\R^n}\prod_p d\xi_p \, e^{- s_{YM} (\th,\xi)}\,\det\left( r^2 (\tilde k(\th) + g_0 \tilde i(\th) \xi + g_0^2 \tilde j(\th) \xi\xi) \right) $$
$$=  r^{-4n|G|}\, r^{2n|G|}\int_{\R^n}\prod_p d\xi_p \, e^{- s_{YM} (\th,\xi)}\,\det\left( \tilde k(\th) + g_0 \tilde i (\th) \xi + g_0^2 \tilde j (\th) \xi\xi \right)\,. $$

Therefore
$$Z_{YM}(L) =  r^{-2n|G|} G_n (\th) \,, $$
since the points $(L_1,...,L_N)$ and $(\tilde L_1,...,\tilde L_N)$ have the same spherical coordinates. Hence the path integral for gravity and YM is given by
\be  Z  = \int_D d^N L \, \m(L) \, e^{\iu S_R^*(L)/l_P^2 } \, r^{-2n|G|} G_n (\th)  \,.\label{ymg}\ee

The integral (\ref{ymg}) is of the same type as in the case of $2|G|$ massles scalar fields, which implies that $Z$ is absolutely convergent for 
\be N > 2n|G| \,,\label{ymcc}\ee 
or 
\be N_1 > 2N_0 |G| \,.\ee

As in the scalar field case, we obtain for a regular triangulation
$${N_1\over 2N_0} \ge {5\over 4}>|G| \,,$$
so that the convergence is guaranteed only for the $U(1)$ gauge group.

Note that the path integral divergencies for the bosonic matter come from the region of arbitrary small edge lengths. One could solve this problem by imposing the minimal edge length restriction, for example
$$|L_\e| \ge l_P \,.$$
However, we will see in section 6, that when combining bosonic and fermionic fields, these divergencies dissapear, provided that one has a sufficient number of fermionic fields.

\section{Dirac spinors}

Let us consider a massive Dirac field $\psi^\a (x)$ on $T(M)$ with the edge lengths $L_\e$. The Dirac action on $M$ can be written as
\be S_D = \int_M \e^{abcd} e_b \wedge e_c \wedge e_d \wedge \bar\psi\left(\iu \g_a \, \textrm{d} + \iu \g_a \, \o - m\,e_a \right)\psi \label{smda}\,,\ee
where $\o = \o^{ab}[\g_a,\g_b]/8$ and $\g_a$ ar the gamma matrices.

The action (\ref{smda}) can be written as
\be S_D = \int_M \e^{abcd} B_{bcd} \wedge J_a \,,\label{31f} \ee
where the 3-forms $B$ are given by
$$ B_{abc} = e_a \wedge e_b \wedge e_c  \,,$$
while the one-forms $J$ are given by
$$ J_a = \bar\psi ( \iu \g_a (\textrm{d} + \o) - me_a ) \psi \,. $$

The action (\ref{31f}) is an integral of a product of a 3-form with a one-form, and any such action on $T(M)$ can be written as
$$\sum_{\t\in T(M)} B_\t J_l \,, $$
where $l$ is the dual edge which corresponds to a tetrahedron $\t$ in $T(M)$ and
$$B_\t = \int_\t B \,,\quad J_l = \int_l J \,.$$

Therefore we obtain
$$ S_D = S_D^{(1)} + S_D^{(2)} + S_D^{(3)} \,,$$
where
\bea S_D^{(1)}(\psi, L) &=& \sum_{\t \in T(M)}V_\t (L) |L_l| \,\e^{abcd} B_{abc}(\t) \, \bar\psi(v) \g_d \, {\psi(v') - \psi(v)\over |L_l|}\nonumber\\
&=&\sum_{\t \in T(M)}V_\t (L) \,\e^{abcd} B_{abc}(\t) \, \bar\psi(v) \g_d \, (\psi(v') - \psi(v))\,,\eea
and 
$$ B_{abc}(\t)  = e_{ai}(\t) e_{bj}(\t) e_{ck}(\t) \e^{ijk} \,,$$
while $v$ and $v'$ are the vertices of the dual edge $l$ which corresponds to a tetrahedron $\t$. The vierbeins on $T(M)$ are defined by
$$ g_{\m\n}(\s) = e_\m^a(\s) e_{\n a}(\s) \,,$$
see \cite{MVb}.

For the second term of the Dirac action one has
\be S^{(2)}_{D}(\psi,L) = \sum_{\t\in T(M)} V_\t |L_l |\,\e^{abcd} B_{abc}(\t)\, \bar\psi(v) \g_d \, \o (l) \psi (v')\,, \ee
while for the mass term one has
\be S^{(3)}_{D}(\psi,L) = m \sum_{\s\in T(M)} V_\s (L) \bar\psi (v) \psi(v) \,.\ee

Note that the Dirac action has been written in terms of the fermionic variables $\psi(v)$, where $v \in T^*(M)$ is a dual vertex. The Dirac action can be also written in terms of the fermionic variables $\psi(\p)$, where $\p$ is a vertex of $T(M)$. In this case we have
$$S_D^{(1)}(\psi^*, L)  = \sum_p B_p J_\e \,, $$
where $p$ is the dual 3-polytope to an edge $\e$. Furthermore
\bea S_D^{(1)}(\psi^*, L) &=& \sum_{p \in T^*(M)}V_p (L) |L_\e| \,\e^{abcd} B_{abc}(p) \, \bar\psi(\p) \g_d \, {\psi(\p') - \psi(\p)\over |L_\e|}\nonumber\\
&=&\sum_{p \in T^*(M)}V_p (L) \,\e^{abcd} B_{abc}(p) \, \bar\psi(\p) \g_d \, (\psi(\p') - \psi(\p)) \nonumber\\
&=&\sum_{\e \in T(M)}V^*_\e (L) \,\e^{abcd} B_{abc}(\e) \, \bar\psi(\p) \g_d \, (\psi(\p') - \psi(\p))  \,,\eea
where $\p$ and $\p'$ are the vertices of the edge $\e$ while
\be B_{abc}(p)  = B_{abc}(\e) = {1\over n_4 (\e)}\sum_{\s; \,\e\in\s } e_{ai}(\s) e_{bj}(\s) e_{ck}(\s) \e^{ijk} \,,\label{bp}\ee
and $n_4(\e)$ is the number of $\s$ that share an edge $\e$ which is dual to $p$.

For the second term of the Dirac action
we obtain
\be S^{(2)}_{D}(\psi^*,L) = \sum_{p\in T^*(M)} V_p |L_\e |\,\e^{abcd} B_{abc}(p)\, \bar\psi(\p) \g_d \, \o (\e) \psi (\p')\,, \ee
where $\p,\p' \in \e$ and
\be\o(\e) = \frac{1}{n_3 (\e)}\,\sum_{\t ;\, \e\in\t }\o(l) \,. \label{oe}\ee
Here $\t$ is a tetrahedron dual to $l$ and $n_3 (\e)$ is the number of $\t$ that share an edge $\e$. For the mass term one has
\be S^{(3)}_{D}(\psi^*,L) = m \sum_{\p\in T(M)} V_\p (L) \bar\psi (\p) \psi(\p) \,,\ee
where $V_\p$ is the volume of the 4-polytope $q$ dual to a vertex $\p$.

In any case, the Dirac action on $T(M)$ can be then written as
$$ S_D (\psi, L) = S_{DK}  + S_{Dm} \,,$$
where
$$S_{DK} = \sum_{\u,\v} \bar\psi_\u C_{\u\v}(L) \psi_\v \,,$$
and
$$S_{Dm} = \sum_{\v} \bar\psi_\v M_{\v}(L) \psi_\v \,,$$
such that $\v =(v,\a) \in T^*_0(M) \times I_D$ or $\v =(\p, \a) \in T_0(M)\times I_D$ where $I_D =\{1,2,3,4\}$.  

The fermionic path integral can be then written as
$$Z_D =  \int \prod_{\v=1}^{4n} d\bar\psi_\v \, d\psi_\v \, e^{\iu (S_{DK} + S_{Dm})/\hbar}\,,$$
where we now assume that $\psi_\v$ and $\bar\psi_\v$ are the generator elements of the corresponding Grassmann algebra, satisfying
$$ \{\psi_\v ,\psi_{\v'} \} = \d_{\v,\v'}\,, \quad  \{\psi_\v ,\bar\psi_{\v'} \} = 0  \,, \quad \{\bar\psi_\v ,\bar\psi_{\v'} \} = \d_{\v,\v'}\,.  $$

The integration on a Grassman algebra is defined as a diferentiation, so that
$$ \int \prod_{\v=1}^F d\bar\psi_\v \, d\psi_\v f(\psi, \bar\psi) = {\pa^{2F} \,f(\psi,\bar\psi)\over \pa\psi_1 \cdots \pa\psi_F \pa\bar\psi_1 \cdots \pa\bar\psi_F} \,,$$
where $F=4n$, so that one obtains
$$Z_D = \const \,\,\det\left(C_{\u\v}(L) + \d_{\u,\v} M_\v (L)\right) \,.$$

Because of the scaling properties
$$C(\l L) = \l^3 C(L) \,, \quad M(\l L) = \l^4 M(L) \,, $$
it is convenient to use the spherical coordinates for a vector $L$, so that we obtain
$$ Z_D = \const \,\det\left(r^3c_{\u\v}(\th) + \d_{\u,\v} r^4 \,m_\v (\th)\right) = r^{12n}\, P (r, \th) \,,$$
where
$$P(r,\th) = \sum_{k=0}^{4n} r^k a_k (\th) \,.$$

Note that in order to calculate $Z_D$ we did not have to perform a Wick rotation on $L$'s as in the case of bosonic matter. But even if we did use the Eucledean path integral
$$\tilde Z_D (\tilde L) =  \int \prod_{\v=1}^{4n} d\bar\psi_\v \, d\psi_\v \, e^{- (\tilde S_{DK} + \tilde S_{Dm})/\hbar}\,,$$
and then used $Z_D(L) = \tilde Z_D (w(L))$, the same results would be obtained.

The convergence of the total path integral
$$ Z = \int_D \prod_{\e =1}^N dL_\e \,\m(L) \, e^{\iu S_R(L)/l_P^2}\, Z_D (L) $$
can be now studied. We use the same approach as in the massless scalar field, so that
$$Z = Z_1 + Z_2 \,,$$
where
$$Z_k = \int_{D_k} d^N L \,\mu(L)\, e^{\iu S^*_R(L)/l_P^2}\,  Z_D(L) \,.$$

Consequently
$$|Z| \le |Z_1|  + |Z_2| \,,$$
and we have 
$$ |Z_1| \le \int_{D_1} d^N L\, |Z_D(L)| = \const\int_0^R r^{N-1 +24n} d r \int_{\O} d^{N-1}\th J_N(\th) |P (r, \th)|$$
$$\le \const \sum_{k=0}^{4n} \int_0^R r^{N-1 +12n + k} d r \int_{\O} d^{N-1}\th J_N(\th) |a_k (\th)|\,,$$
and all these integrals are convergent.

For $Z_2$ we have
$$ |Z_2| \le  \int_{D_2} d^N L\, |Z_D(L)|\prod_\e {l_0^{2p} \over( l_0^2 + L_\e^2)^p} \,,$$
so that
$$|Z_2| \le \const\int_R^\infty r^{-(2p-1)N-1 +12n} d r \int_{\O} d^{N-1}\th J_N(\th)\,h(\th, l_0/r)\, |P (r, \th)|\,,$$
where
\be \prod_\e  {l_0^{2p} \over (l_0^2 + L_\e^2)^p} =r^{-2Np}\prod_\e  {l_0^{2p} \over (\frac{l_0^2}{r^2} + f_\e^2(\th))^p} = r^{-2Np} h_N (\th, l_0/r) \,.\label{dfh}\ee

Then
$$|Z_2| \le \const \int_R^\infty r^{-(2p-1)N-1 +12n} d r \int_{\O} d^{N-1}\th J_N(\th) \,h_N (\th, l_0/r)\, A_n r^{4n}\,,$$
since for a sufficiently large positive constant $A_n$, we have
$$ |P(r,\th)| \le A_n r^{4n} $$
for $r \ge R$. Consequently
$$|Z_2| \le \const \int_R^\infty r^{-(2p-1)N-1 +16n} d r \int_{\O} d^{N-1}\th J_N(\th) \,h_N (\th, l_0/r)\,,$$
so that the absolute convergence requires 
\be (2p-1)N > 16n \,, \label{fcc}\ee
or
\be (2p-1)N_1 > 16N_4 \,.\ee

In the case that we use for the fermionic DOF the verices of $T(M)$, then $n=N_0$ and $N=N_1$, so that
\be (2p-1)N_1 > 16 k N_0 \,,\label{fcc}\ee
where $k$ is the number of the Dirac fermions.

Note that the PI convergence criterion for fermions (\ref{fcc}) now comes from the integration over the region of large edge lengths, in contrast to bosons, where such a criterion comes from the integral over the small edge lengths. 

Also note that (\ref{fcc}) will be satisfied for 
\be p-1/2 > 16k/5 \,,\label{dsc}\ee
since $N_1/N_0 \ge 5/2$ for a regular triangulation. Hence for a sufficiently big $p$ we will insure the convergence of the fermionic PLQG path integral. 

In the case of chiral Dirac spinors, also known as the Weyl spinors, which are relevant for the Standard Model,
$$ \psi_L = {1-\g_5\over 2} \psi\,, \quad \psi_R = {1+\g_5\over 2} \psi\,.$$
In this case we have only 2 independent components per spinor, so that the convergence condition (\ref{dsc}) becomes
\be p-1/2 > 8k/5 \,.\ee

\section{Standard Model}

The action for the Standard Model on $M$ contains massles scalar fields, YM gauge fields and chiral Dirac spinors. We can write it as
$$S_0 = S_H + S_{YM}  + S_f  + S_{Y}  = \int_M d^4 x \,\sqrt{g}\left(\cl_H + \cl_{YM} + \cl_f + \cl_Y \right) \,,$$
where 
$$ \cl_H = \frac{1}{2} D^\m \f^\dag  D_\m \f - \l_0^2 (\f^\dag \f - \f_0^2)^2 \,,$$
$$ \cl_{YM} = -\frac{1}{4}\, Tr\left(F^{\m\n} F_{\m\n} \right) \,,$$
$$ \cl_f = \sum_{k=1}^{48} \e^{abcd} e_b \wedge e_c \wedge e_d \, \bar\psi_k \left(\iu \g_a ( \textrm{d} + \iu \o +  \iu g_0 A) \right)\psi_k \,, $$
$$\cl_Y = \sum_{i,j} Y_{ij}\,\langle \bar\psi_i \psi_j \f \rangle\,,$$
and
$$D_\m \f = (\pa_\m  + \iu \,(g_0 A)_\m )\f  \,.$$

Here the gauge group $G$ is given by $U(1) \times SU(2) \times SU(3)$ and
$$ \f = \t_a \f^a \,, \quad A = T_a A^a \,,$$
where $\t_a$ are the Lie algebra generators of $G$ in the Higgs field representation, while $T_a$ are the corresponding matrices in the adjoint representation of $G$. We also use a short-hand 
$$g_0 A = \sum_{l=1}^3 g_{0l}A_l \,,$$
where $A_1$, $A_2$ and $A_3$ belong to the Lie algebras of $U(1)$, $SU(2)$ and $SU(3)$, respectively and $g_{0l}$ are the corresponding coupling constants.

A fermion $\psi_k$ is a Weyl spinor $\psi_R$ or $\psi_L$, can be a lepton or a quark, and belongs to an appropriate representation of $G$. There are 3 generations of fermions which gives 48 Weyl spinors. We use  $\langle ...\rangle$ in $\cl_Y$  to denote the corresponding group invariant.

On $T(M)$ we will use as the DOF the values of these fields at the vertices of $T(M)$, so that
$$ \tilde S_H = \sum_\s V_\s (L) \, s_{HK} + \sum_\p V_\p^* (L) \, s_{HP} \,,$$
where
$$ s_{HK} =g^{\m\n}_\s  \left(\frac{\f(\p_\m) - \f (\p_0)}{|L_{0\m}|} + \iu g_0 A_\m (\p_0) \f_{\p_0} \right)^\dag \left(\frac{\f(\p_\n) - \f (\p_0)}{|L_{0\n}|} + \iu g_0 A_\n (\p_0) \f_{\p_0} \right)$$
and
$$ s_{HP} =    \l_0^2  \left(\f^\dag (\p)\f (\p) - \f_0^2 \right)^2 \,.$$

Note that $\tilde S_H$ is a Euclidean action, so that $\tilde S_H >0$, which we need in order to define the matter path integral. The euclidean PL geometry will be defined by a Wick rotation $L_\e \to \tilde L_\e = |L_\e|$, see section 3. The same applies to the YM action, which will be given by (\ref{eym}) plus the gauge-fixing term (\ref{egf}) and the ghost action (\ref{egh}) .

The fermion action on $T(M)$ is given by
$$ \tilde S_f = \sum_{\e} V_\e^* ( L) \, s_{f} +\sum_{\p} V_\p^* ( L) \, \,s_{YMf} \,,$$
where
$$ s_{f} = \sum_{i} \e^{abcd} \, B_{abc}(p) \,\bar\psi_i(\p)\,\iu \g_d\,(| L_\e| \,\iu \o_\e (L)\,\psi_i(\p') + \psi_i(\p') - \psi_i(\p)) \,,$$
and 
\be  s_{YMf} = \sum_{i}  \bar\psi_i(\p) \, g_0 \g^\m (\p) A_{\m}(\p)\, \psi_i(\p) \,.\label{plymf}\ee

In (\ref{plymf}) we used
$$\g^\m (\p) = e^\m_a (\p) \g^a \,,$$
where
$$e^\m_a (\p) ={1\over n_\s(\p)} \sum_{\s ;\, \p\in \s} e^\m_a (\s) \,.$$

The Yukawa action on $T(M)$ is given by
$$\tilde S_Y = \sum_\p V_\p^* (L)\,  s_Y \,, $$
where
$$s_Y = \sum_{i,j} Y_{ij} \langle\bar\psi_i(\p) \psi_j (\p)\f (\p) \rangle\,.$$

Note that
$$ V_\e^*(L) = V_p (L) \,,\quad V_\p^* (L) = V_{q}(L) \,,$$
where $p$ is the 3-polytope dual to an edge $\e$ and $q$ is the 4-polytope dual to a vertex $\p$ of $T(M)$.
Also note that there are no mass terms for the fermions, since they are generated through the Yukawa couplings. 

If we pass to the spherical coordinates for the edge lengths, then we can represent schematically the gauge-fixed SM action on $T(M)$ as
$$ S = S_1 + S_2 +  \tilde S_2 + S_3 \,, $$
where
\bea S_1 &=& r^3\langle  \bar\psi\psi \rangle + r^4\langle \bar\psi\psi A \rangle + r^4\langle \bar\psi\psi \f \rangle \nonumber\\
S_2 &=&  r^4 \langle ( Ar^{-1} + g_0 A^2 )^2\rangle   \nonumber\\
\tilde S_2 &=& r^3 \langle \bar c ( r^{-1} + g_0 A c )\rangle \nonumber\\
S_3 &=& r^4 \langle ( \f r^{-1} + g_0 A\f )^2\rangle + \l_0^2 r^4 \langle (\f^2 - \f_0^2)^2\rangle  \,,\eea
where a bracket $\langle XY\cdots \rangle$ represents
$$ \sum_{\a,\b,...} c^{\a\b ...}(\th) X_\a Y_\b \cdots \,. $$

Then we have
$$ \int d^{nn_f}\bar\psi \, d^{nn_f}\psi \, e^{-S} = e^{-S_2 -\tilde S_2 - S_3} \,\det \left(r^3\langle 1 + rA + r\f\rangle_{kl} \right)  $$
$$ = r^{3n\cdot n_f}\,e^{-S_2 - \tilde S_2 - S_3} \,\det\left( \langle 1 + rA + r\f\rangle_{kl} \right) \,,$$
where $n_f$ is the number of complex components of the fermionic fields of the Standard Model. Since the SM has 16 Weyl spinors in a fermionic generation and there are 3 generations, then
\be n_f = 2\cdot 16 \cdot 3 = 96 \,. \ee

Therefore the matrix $\langle 1 + rA + r\f\rangle$ has the size $96n \times 96n$, and an entry $\langle 1 \rangle_{kl}$ denotes a function of $\th$.

Integration of the YM ghosts gives
$$ r^{3n_f n + 2|G|n}\,e^{-S_2 - S_3} \,\det \left( \langle 1 + rA + r\f\rangle_{kl}\right)   \,\det \left(\langle 1 + rA \rangle_{rs}\right)\,,$$
where $\langle 1 + rA \rangle_{rs}$ is a $2|G|n \times 2|G|n$ matrix.

Now we make the change of variables
$$ \xi = r A \,,\quad \chi = r\f \,, $$
so that we obtain 
$$\tilde Z_m =  r^{(3n_f - 2|G| - 4)n  }\,\int d^{4n} \chi \,d^{4|G|n} \xi \, e^{-S_2 - S_3} \,\det \langle 1 + \xi + \chi \rangle   \,\det \langle 1 + \xi \rangle\,,$$
where
$$S_2 + S_3 = \langle (\xi + g_0 \xi^2 )^2\rangle + \langle (\chi + g_0 \xi\chi)^2 \rangle + \l_0^2 \langle (\chi^2 - r^2\f_0^2)^2\rangle \,.$$

Since
$$S_2 + S_3 > \langle (\xi + g_0 \xi^2 )^2\rangle + \langle (\chi + g_0 \xi\chi)^2 \rangle \equiv s(\theta, \xi, \chi) \,,$$
then
$$|\tilde Z_m(\tilde L)| <  r^{(3n_f - 2|G| - 4)n  }\,\int d^{4n} \chi \,d^{4|G|n} \xi \, e^{-s(\th,\xi,\chi)} \,\det \langle 1 + \xi + \chi \rangle   \,\det \langle 1 + \xi \rangle\,,$$
$$ \equiv  r^{(3n_f - 2|G| - 4)n} \, F_n (\th) \,.$$

Given that
\be Z = \int_D d^N L \, \m(L) \, e^{\iu S_R^*(L)/l_P^2} Z_m(L) \,, \label{pigm}\ee
where 
$$Z_m(L) = \tilde Z_m (w(L)) $$
then
$$|Z| \le \int_D d^N L \, \m(L) \,  |Z_m(L)| \,< \,\int_D d^N L \, \m(L) \, r^{(3n_f - 2|G| - 4)n} \, F_n (\th) \equiv I \,. $$

Since 
$$I = I_1 + I_2 $$
where
$$I_1 = \int_0^R r^{N-1} dr \int_\O d^{N-1}\th \, J_N(\th) \, \m(r,\th)   r^{(3n_f - 2|G| - 4)n} \, F_n (\th) $$
and
$$I_2 = \int_R^\infty r^{N-1} dr \int_\O d^{N-1}\th \, J_N(\th) \, \m(r,\th)   r^{(3n_f - 2|G| - 4)n} \, F_n (\th) \,,$$
then the convergence of $I_1$ is determined by the integral for small $R$, while the convergence of $I_2$ is determined by the integral for large $R$.

Since
$$\m(r,\th) \approx 1 $$
for $r\to 0$, while
$$\m(r,\th) \approx r^{-2pN}h_N(\th, 0) $$
for $r\to\infty$, see eq. (\ref{dfh}), then
$$ I_1 \approx \int_0^R r^{cn + N-1} dr \int_\O d^{N-1}\th \, J_N(\th) \, F_n (\th) \,,$$
for small $R$, so that it is convergent when
\be cn + N > 0 \,.\label{ccsr}\ee

Since 
\be c = 3n_f -2|G| - 4 = 3\cdot 96 - 2\cdot (1+3+8) - 4 = 230  > 0 \,,\label{smc}\ee 
then $I_1$ will be convergent. When $R\to \infty$, we obtain
$$ I_2 \approx \int_R^\infty r^{cn -(2p-1) N-1} dr \int_\O d^{N-1}\th \, h_N(\th,0)\, J_N(\th) \, F_n (\th) \,,$$
which is convergent for
$$cn - (2p-1)N < 0 \,,$$
or
\be {c \over (2p-1)} < {N\over n} \,.\label{cclr}\ee

Therefore the PI for GR and SM will be convergent for triangulations which satisfy (\ref{cclr}). For a regular triangulation we have
$$ {N \over n} = {N_1 \over N_0} \ge {N_1^* \over N^*_0} = \frac{5}{2} \,.$$
Hence if we choose $p$ such that
$$ {c \over (2p-1)} < \frac{5}{2} \le {N\over n} \,,$$
then the path integral will be absolutely convergent. Since in the case of the SM  $c=230$, see (\ref{smc}), we then obtain
\be p - \frac{1}{2} > 46 \,. \label{smp}\ee

Therefore we can insure the absolute convergence of the path integral $Z$ for GR and SM if we choose the parameter $p$ to be sufficiently large, i.e. $p > 46,5$. 

\section{Conclusions}

The potential divergencies of a path integral for gravity and bosonic matter on a PL spacetime $T(M)$ come from the region of small edge lengths, i.e. when $r \to 0$. In this case we showed that the path integral is convergent when there is a small number of fields and $p > 1/2$.  In the case of fermionic matter, the potential divergencies come from the region of  large edge lengths, i.e. when $r \to \infty$. We then showed that the path integral is convergent for sufficiently large values of the parameter $p$. 

If we combine bosonic and fermionic matter, as in the case of the SM, the short distance divergencies of the bosonic fields are cured by the fermionic fields, since in the SM the triple number of fermionic fields components is greater than the number of bosonic fields components, see (\ref{smc}). The long-distance divergencies of the fermionic fields are cured by a choice of a sufficiently large value for $p$, which in the SM case is given by $p > 46,5$.

The path integral for GR coupled to the SM may be convergent for a lower value of $p$ than $46,5$, but in order to explore this, one needs to find more powerfull techniques than the ones used here. However, at this stage it is more important to show that an interval of values of $p$ exists which ensures the convergence of the path integral. Consequently, the PLQG theory presented here is the first example of a mathematically complete QG theory. It is a complete QG theory in the sense that it is finite, and the corresponding effective action has the correct classical limit and there is a semiclassical expansion that corresponds to a perturbative QFT quantization of GR coupled to the SM. This semiclassical expansion is given by the usual QFT perturbative expansion around a background metric, with a physical cutoff determined by the average edge length in the local patch $T(B_3 \times I)$.

Note that the string theory is also a finite QG theory, but the classical limit is a 10-dimensional supergravity theory, so that one has to show that there is a compactification and a symmetry breaking such that we see only the SM particles at the LHC experiments. In the case of LQG, the spin foam models can be also made finite on a PL spacetime \cite{MV2}. However, the classical limit of the spin foam effective action is not the Regge action, but it is given by the area-Regge action \cite{M3}. Furthermore, it is not known how to couple the fermionic matter to a SF model.

The question whether a PL quantum gravity (PLQG) theory is realized in Nature or not, can be decided by experiments and observations. At the moment there are no QG phenomena which can be studied in an experiment, but it is concievable that in a not so distant future, an elementary particle collider of the energies of 100 TeV or 1000 TeV will be built.  One will then probe the distancies of order $10^{-22}$ m, which could be close enough to the average edge length in the corresponding local patch $T(B_3 \times I)$. In this case one could detect the deviations from the smooth spacetime QFT scattering results. 

Another type of tests for a PLQG theory would be a comparison of the present cosmological observations to the corresponding theoretical predictions. For example, in \cite{MV, MVr} it was shown that the observed value of the cosmological constant (CC) belongs to the interval of possible values for the CC in a PLQG theory. Note that in the proof given in \cite{MV, MVr}, the fundamental assumption was the finitenness of the path integral for gravity and matter. Hence our main result closes the gap in the proof that the PLQG CC spectrum contains the observed value of the CC. 

Note that in the case of string theory, proving that the observed CC value belongs to the CC spectrum  is a complicated problem, because a natural value for the string CC is negative and the spectrum is discrete \cite{P}. That is why a complete proof in string theory has not been achieved yet, and only the plausibility arguments were given. 

As far as the question why the CC takes the value that we observe, that is a more difficult question, and answering this question would require a wider theoretical framework than the one used in this paper. Note that the formulation of PLQG used in this paper has three free parameters $(L_0, l_0, p)$, but one can also add the fourth parameter $\L_c$, which corresponds to the classical CC term in the action. The demonstration that a PLQG theory with the SM matter is finite for $p > 46,5$, gives us the first example of a simple and mathematically complete theory of quantum gravity.

\bigskip
\bigskip
\noindent{\Large\bf Acknowledgments}

Partially supported by the Science Fund of the Republic of Serbia, grant 7745968, Quantum Gravity from Higher Gauge Theory 2021, QGHG-2021.

\end{document}